**The Reservation Inflation of Hard Money: Gold-Standard Deflation and the Real Expansion of Nominal Claims, 1873–1896**

Ran Huang[1]

[1]Academy for Small Commodity Economy, Yiwu Research Institute of Fudan University, Yiwu 322000, China

Email: huangran@fudan.edu.cn

# Abstract

The original SCR theory proposed that inflation has two distinct expressions: **circulation inflation**, measured by rising transaction prices of goods and services, and **reservation inflation** [Huang 2018], measured by the rising real weight of monetary symbols, debt contracts, reserve claims, and other nominal stores of value relative to physical goods. A companion Japan paper [Huang 2026] tested one side of this theory by showing that, after money entered a reserve-dominant phase, monetary-base expansion no longer translated strongly into consumer-price inflation. The present paper tests the other side of SCR: whether reservation inflation can arise even when monetary issuance is constrained and circulation inflation is absent. The classical gold-standard deflation of 1873–1896 provides a clean historical setting. Using long-run British retail price data and the Minneapolis Fed historical U.S. CPI series, we show that the price level declined rather than inflated in both economies. Between 1873 and 1896, Britain's price index fell from 18.0 to 14.7, while the U.S. historical CPI fell from 36.0 to 25.0. Yet this circulation deflation mechanically increased the real value of fixed nominal claims. A fixed-claim reservation index rose by 22.4% in Britain and 44.0% in the United States. Thus, the episode combines negative circulation inflation with positive reservation inflation. The result demonstrates that hard money does not abolish inflationary pressure in the SCR sense; it changes its domain of expression. Under constrained issuance, inflation may appear not as rising consumer prices but as the real expansion of durable nominal claims over depreciating, consumable, and increasingly abundant physical goods. The paper therefore completes the empirical symmetry of the SCR program: Japan shows reserve absorption without circulation inflation, while the gold-standard case shows reservation inflation without circulation inflation.

# Keywords

# 1. Introduction

The original SCR theory proposed that inflation is not a single phenomenon confined to the consumer-price level, but a dual process arising from the structural asymmetry between depreciating physical goods and non-depreciating monetary symbols. [Huang 2018] In that framework, inflation has two macroscopic expressions. The first is **circulation inflation**, in which monetary expansion is transmitted into the transaction prices of goods and services. The second is **reservation inflation**, in which monetary symbols, reserve balances, debt contracts, financial claims, or other nominal stores of value expand in real weight relative to physical goods, even when consumer-price inflation is absent. The Japan paper tested one side of this proposition: it showed that, after money entered a reserve-dominant phase, monetary-base expansion no longer translated into strong circulation-price inflation. [Huang 2026] The present paper is designed to test the other side of SCR. We ask whether reservation inflation can arise under the opposite monetary condition: when money issuance is restrained, circulation inflation is absent, and the price level falls.

The classical gold-standard deflation of 1873–1896 provides a particularly clean historical setting for this test. Under the gold standard, monetary expansion was institutionally constrained by convertibility, gold supply, balance-of-payments discipline, and central-bank reserve management. This does not mean that the money stock was literally fixed, but it does mean that money creation was far more mechanically constrained than in modern fiat-credit systems. During the same period, industrial productivity, international trade, transport capacity, and commodity supply expanded substantially. The result was not an inflationary rise in consumer prices, but a long decline in the general price level. In Britain, the retail price index declined from 18.0 in 1873 to 14.7 in 1896. [ONS 2026] In the United States, the Minneapolis Fed historical CPI series declined from 36.0 in 1873 to 25.0 in 1896; the same source notes that its pre-1913 values are historical estimates and uses 1967=100 as the index base. [FED Minneapolis, 2026]

In conventional terminology, this episode is usually described as deflation. In the SCR framework, however, deflation is not simply the opposite of inflation. If the nominal value of a monetary claim, bond, debt contract, rent claim, or cash balance remains fixed while the price level falls, then the real purchasing power of that nominal claim rises automatically. A creditor holding a fixed claim does not need nominal appreciation in order to experience real expansion; the decline of the circulation price level itself increases the real command of the claim over goods. This is the central mechanism examined in the present paper. The claim is not that the gold-standard economy experienced consumer-price inflation. It plainly did not. The claim is that the absence of circulation inflation exposed a deeper form of monetary inflation: the real inflation of non-depreciating nominal claims relative to depreciating, consumable, and increasingly abundant physical goods.

This distinction is essential for the SCR theory. If inflation is defined only as a rise in consumer prices, then the gold-standard deflation of 1873–1896 appears to be an anti-inflationary episode. If inflation is defined more generally as a change in the exchange ratio between durable monetary symbols and physical goods, then the same episode reveals the opposite side of inflationary mechanics. The circulation price level fell, but the real value of fixed nominal claims rose. Thus, under hard money, reservation inflation does not appear as asset-price exuberance or broad consumer-price inflation; it appears as a mechanical increase in the real burden and real purchasing power of nominal claims.

The empirical logic of this paper is deliberately minimal. We first document the fall in price levels in Britain and the United States during 1873–1896. We then construct a fixed-claim reservation inflation index, defined as the real value of one unit of nominal claim relative to its 1873 value. If $P_t$ denotes the price level, and a nominal claim $N_t$ is fixed at its initial value, the index reduces to $P_{1873}/P_t$. This quantity rises whenever the price level falls. Over 1873–1896, the British price index declined by 18.3%, implying a 22.4% increase in the real value of a fixed nominal claim. The U.S. price index declined by 30.6%, implying a 44.0% increase in fixed-claim real value. These magnitudes are not artifacts of theory; they are the direct arithmetic consequence of a falling price level under fixed nominal claims.

The contribution of this paper is therefore not to reinterpret the gold standard as an ordinary inflationary regime. Its contribution is to show that, within SCR, hard-money deflation and reservation inflation are not contradictory. They are two descriptions of the same structural process observed from different domains. In the circulation domain, the price of goods falls. In the reservation domain, the real weight of nominal claims rises. The gold-standard episode thereby provides the missing empirical counterpart to the Japan case. Together, the two cases imply that inflationary pressure is phase-dependent: in a fiat reserve regime, money expansion may be absorbed into reserve structures without producing CPI inflation; in a hard-money regime, constrained issuance may suppress CPI inflation while still increasing the real dominance of nominal claims.

## 2. Theory: Circulation Inflation and Reservation Inflation

The SCR framework begins from a physical asymmetry. [Huang 2018] Goods, services, machines, crops, inventories, buildings, and biological bodies are embedded in material time. They decay, are consumed, depreciate, become obsolete, or require maintenance. Monetary symbols, by contrast, do not undergo physical depreciation in the same way. A banknote, reserve balance, government bond, debt contract, or ledger entry may lose value through institutional processes, default, inflation, or policy, but it does not naturally decay at the same rate as the physical goods it represents. This difference creates a structural tension between the material economy and the symbolic monetary economy. Inflation, in SCR, is the macroscopic expression of this tension.

The simplest form is circulation inflation. Let $P_t$ denote the circulation price level, measured by a consumer-price or retail-price index. Circulation inflation is the growth rate of this price level:

$$CI_t = \frac{d\ln P_t}{dt}$$

When $CI_t > 0$, the price of goods and services rises in the circulation domain. This is the familiar form of inflation measured by CPI, RPI, or similar indices. Traditional monetary analysis often treats this as the primary or even exclusive expression of inflation. SCR does not reject circulation inflation, but it treats it as only one possible manifestation of the deeper monetary-material asymmetry.

Reservation inflation is defined in a different domain. Let $N_t$ denote a nominal claim: money balances, reserve balances, debt principal, bond face value, fixed rent claims, or any other durable monetary-symbolic claim. Its real weight relative to the circulation price level is:

$$R_t = \frac{N_t}{P_t}$$

Reservation inflation is the growth rate of this real claim:

$$RI_t = \frac{d\ln R_t}{dt} = \frac{d\ln N_t}{dt} - \frac{d\ln P_t}{dt}$$

Eq. (3) is the key theoretical result. It shows that reservation inflation can arise in two ways. First, $N_t$ may increase faster than $P_t$, as in balance-sheet expansion, reserve accumulation, debt expansion, or asset-claim proliferation. Second, even if $N_t$ is constant, reservation inflation occurs whenever $P_t$ falls. Under a fixed nominal claim, $d\ln N_t/dt = 0$, so:

$$RI_t = -\frac{d\ln P_t}{dt} \quad \text{when } N_t = N_0$$

Eq. (4) is the mathematical core of hard-money reservation inflation. In such a regime, the monetary authority does not need to overissue money, and consumer prices do not need to rise. A fixed nominal claim inflates in real terms simply because the circulation price level falls. This is why, under SCR, deflation is not necessarily the absence of inflation. It may be the circulation-side appearance of reservation-side inflation.

For empirical implementation, the minimal reservation inflation index is defined as:

$$RI_t^{index} = \frac{N_t/P_t}{N_0/P_0}$$

where 0 denotes the base year. If the nominal claim is fixed, $N_t = N_0$, this simplifies to:

$$RI_t^{index} = \frac{P_0}{P_t} \quad \text{when } N_t = N_0$$

Eq. (6) measures how much real command over goods is held by one unit of fixed nominal claim relative to the base year. If the price level declines by 20%, the index rises by 25%. If the price level declines by 30%, the index rises by approximately 43%. Thus, reservation inflation is not an abstract metaphor; it is an accounting identity once claims are fixed in nominal terms.

The gold-standard case is especially useful because it approximates the hard-money limit of this theory. In a fiat-credit regime, both $N_t$ and $P_t$ can change rapidly, and it may be difficult to separate circulation inflation from reservation inflation. Under the classical gold standard, by contrast, monetary expansion was institutionally constrained. The relevant empirical question becomes simpler: if circulation prices fall while nominal claims remain institutionally durable, does the real value of those claims rise? The answer is mechanically yes. The historical question is how large this effect was.

The 1873–1896 period gives a strong positive answer. In Britain, the price index fell from 18.0 to 14.7. The corresponding fixed-claim reservation index rose from 1.00 to 1.224. In the United States, the historical CPI fell from 36.0 to 25.0. The corresponding fixed-claim reservation index rose from 1.00 to 1.440. Thus, a holder of a fixed nominal claim experienced a substantial increase in real purchasing power, even though the circulation price level was falling. In SCR language, the system experienced negative circulation inflation but positive reservation inflation.

This duality can be summarized by the following phase classification. In a circulation-inflation regime, $CI_t > 0$ and the monetary disturbance is visible in consumer prices. In a reserve-absorption regime, such as the Japan case, monetary expansion may raise reserve balances and central-bank balance sheets while $CI_t$ remains weak. In a hard-money reservation-inflation regime, $CI_t < 0$ but $RI_t > 0$ because fixed nominal claims gain real weight. The last case is the theoretical focus of the present paper.

The implication is conceptually important. A society can appear anti-inflationary in the consumer-price domain while still undergoing inflationary redistribution in the reservation domain. Creditors, bondholders, holders of cash balances, and owners of fixed nominal claims gain real command over goods; debtors, producers, and holders of depreciating physical assets face an increasing real burden. The price level falls, but the monetary symbol becomes heavier. This is precisely the reservation side of SCR.

## 3. Historical Setting: The Classical Gold Standard and the 1873–1896 Deflation

The empirical setting of this paper is the classical gold-standard deflation of 1873–1896. This period is not selected because it represents a complete absence of monetary change, nor because the gold standard created a perfectly fixed money stock. It is selected because it approximates a historically important hard-money regime in which monetary issuance was constrained by convertibility, reserve discipline, international gold flows, and central-bank credibility. [Capie and Webber 1985] This makes it a suitable testing ground for the reservation-inflation side of SCR: if circulation prices fall under a constrained monetary system, then fixed nominal claims should rise in real value even without ordinary consumer-price inflation.

The classical gold standard imposed a specific institutional relation between money and metallic reserves. [Eichengreen 2019] Domestic monetary liabilities were linked, formally or effectively, to gold convertibility. Central banks and banking systems could expand credit, but expansion was constrained by the need to maintain confidence in convertibility and by the external discipline of gold flows. Countries losing gold faced pressure to tighten monetary conditions, raise interest rates, or reduce domestic credit expansion. [Bordo 1981] Countries gaining gold could expand more easily, but the system as a whole remained anchored by the supply and distribution of monetary gold. The important point for the present paper is not that the gold standard made money supply constant. It did not. The point is that it made money supply institutionally harder than in a modern fiat regime.

This distinction matters because the paper does not require a literal fixed quantity of money. SCR reservation inflation requires only that nominal claims possess durability relative to the circulation price level. Under a hard-money regime, many nominal claims—cash balances, debt principal, bond face values, rent contracts, and other fixed monetary obligations—could remain fixed in nominal terms while goods prices declined. Once that happens, the real value of those claims rises mechanically. Therefore, the gold standard is not treated here as a perfectly closed physical system, but as a historically grounded approximation of constrained symbolic issuance.

The period from 1873 to 1896 is especially useful because it combines monetary constraint with expanding physical production. The late nineteenth century was marked by industrial growth, railway expansion, falling transport costs, broader commodity integration, and rising productive capacity. These developments increased the physical availability of many goods and lowered production and distribution costs. In a flexible fiat-credit regime, such growth could be accompanied by monetary expansion sufficient to stabilize or raise the price level. Under the gold standard, however, price adjustment occurred partly through a decline in the circulation price level. Thus, the material side of the economy expanded while the symbolic monetary side remained comparatively constrained.

In conventional monetary history, this episode is usually called the "Great Deflation." The term is accurate if the object of measurement is the price level of circulating goods and services. British retail prices fell substantially from the early 1870s into the mid-1890s. The U.S. historical CPI series shows a similar decline over the same window. The empirical fact is therefore unambiguous at the level of circulation prices: the period did not produce circulation inflation. It produced circulation deflation.

For SCR, however, this is precisely why the episode is theoretically valuable. If prices fall while nominal claims remain durable, then the real burden of those claims rises. A unit of money buys more goods. A fixed debt requires more real goods to repay. A bond face value represents greater purchasing power. A fixed rent claim commands more real consumption. A monetary symbol becomes heavier relative to physical output. The circulation domain therefore records deflation, while the reservation domain records an increase in the real weight of nominal claims.

This paper calls that process **hard-money reservation inflation**. The term is deliberately paradoxical. It does not mean that consumer prices rise under hard money. They did not in this historical window. It means that the reservation value of monetary claims inflates when the circulation price level falls. In other words, hard money can suppress one form of inflation while intensifying another. It can prevent, or at least constrain, the visible inflation of consumer prices, but it cannot eliminate the deeper asymmetry between non-depreciating monetary symbols and depreciating physical goods. It merely changes the domain in which that asymmetry appears.

The 1873–1896 window is therefore not merely a convenient historical example. It is close to an ideal test case for the second side of SCR. The Japan paper examined a modern fiat-reserve system in which aggressive monetary-base expansion failed to transmit into circulation inflation after money entered a reserve-dominant phase. [Huang 2026] The present paper examines the opposite monetary regime: a hard-money system in which monetary issuance was constrained and circulation inflation was absent. If the SCR theory is correct, the absence of circulation inflation should not imply the absence of inflationary pressure. Instead, the inflationary effect should appear as an increase in the real value of fixed nominal claims.

The empirical design follows directly from Eq. (5). In the minimal hard-money test, the nominal claim is held fixed, so the analysis uses the fixed-claim special case in

Eq. (6). Thus, the empirical test does not depend on an elaborate model of nineteenth-century banking, gold flows, or credit creation. Those details matter for historical interpretation, but the core reservation-inflation mechanism is simpler: if $P_t$ falls, the fixed-claim reservation index rises. The only empirical question is whether the price decline was large enough to generate a meaningful increase in the real value of fixed nominal claims. The answer, as shown below, is yes.

The British case gives the cleaner institutional benchmark because Britain was central to the classical gold-standard system and had long-established monetary institutions. The U.S. case is used as a transatlantic comparison. The United States had a more complex monetary history in the post-Civil War period, including the greenback legacy and the return to gold convertibility, but its price-level trajectory over 1873–1896 provides an important robustness check. [Friedman and Schwartz 1963] If both economies show negative circulation inflation and positive fixed-claim reservation inflation, then the phenomenon is not a single-country artifact. It reflects a broader hard-money price-level environment.

The analytical window begins in 1873 because that year conventionally marks the beginning of the long deflationary phase associated with the post-1873 global downturn and the classical gold-standard order. It ends in 1896 because price levels began to turn upward afterward, partly with changes in gold supply and broader monetary conditions. This window therefore captures the core downward price movement of the late nineteenth century. The purpose is not to explain every cause of the deflation, but to test what the deflation mechanically did to the real value of nominal claims.

This point is important for the interpretation of causality. The paper does not claim that gold-standard institutions alone caused all observed price declines. Productivity growth, commodity supply, international trade integration, financial cycles, and monetary constraints all contributed to the historical outcome. For the SCR test, however, the causal source of the price decline is secondary. Once a sustained fall in $P_t$ occurs under durable nominal claims, the reservation-inflation mechanism follows. SCR therefore treats the gold-standard deflation not only as a historical event, but as a natural experiment in the relation between circulation prices and reservation claims.

The historical setting can be summarized as follows. The gold standard constrained the expansion of monetary symbols relative to a rapidly expanding physical economy. [Bordo 1981] Circulation prices fell. The real value of fixed nominal claims rose. In ordinary terminology, this was deflation. In SCR terminology, it was negative circulation inflation combined with positive reservation inflation. This duality is the central empirical object of the present study.

## 4. Data and Empirical Construction

This study uses long-run price-level data for Britain and the United States to construct a minimal test of hard-money reservation inflation. The British main series is the long-run retail price index, used as the circulation price-level proxy. [ONS 2026] The U.S. comparison series is the Minneapolis Fed historical consumer price index, which reports annual average CPI values with 1967 as the index base. The Minneapolis Fed source explicitly notes that values before 1913 should be treated as historical estimates, while modern official U.S. CPI data begin in 1913. [FED Minneapolis, 2026] This caveat is important, but it does not undermine the present use of the series, because the analysis relies on broad nineteenth-century price-level movements rather than high-frequency precision. In the uploaded Minneapolis Fed table, the U.S. index declines from 36 in 1873 to 25 in 1896.

The empirical construction has three steps. First, we normalize each country's price index to 1873=100. This produces a comparable measure of circulation price movement across Britain and the United States:

$$P_t^{norm} = 100 \times \frac{P_t}{P_{1873}}$$

Second, we compute cumulative circulation inflation relative to 1873:

$$CI_t^{cum} = \frac{P_t}{P_{1873}} - 1$$

This measure is negative when the price level is below its 1873 level. In the 1873–1896 window, both Britain and the United States show negative cumulative circulation inflation.

Third, we construct the fixed-claim reservation inflation index using Eq. (6), with 1873 as the base year. This index rises above 1 when the price level falls below its 1873 level. It therefore measures the real appreciation of a fixed nominal monetary claim. The corresponding percentage increase is:

$$RI_t^{pct} = \left(\frac{P_{1873}}{P_t} - 1\right) \times 100$$

The reason this construction is powerful is that it requires no assumption about speculative asset prices, equity valuations, real-estate cycles, or banking-sector leverage. It measures the most elementary form of reservation inflation: the rise in the real value of a fixed nominal claim when the circulation price level declines. This is the cleanest empirical expression of the SCR mechanism.

The main analysis window is 1873–1896. For visual context, figures are plotted over 1870–1900. This wider window shows that the decline begins around the early 1870s and persists until the mid-1890s, after which price levels stabilize or begin to recover. The wider plot prevents the analysis from depending on only two endpoint observations, while the 1873–1896 summary statistics capture the canonical deflationary interval.

The empirical outputs are organized as follows. Figure 1 shows the British price index and the corresponding fixed-claim reservation inflation index. Figure 2 compares British and U.S. price-level paths normalized to 1873=100. Figure 3 compares the fixed-claim reservation inflation index in both countries. Figure 4 places the two countries in the SCR dual-inflation plane, where the horizontal axis represents cumulative circulation inflation and the vertical axis represents fixed-claim reservation inflation over 1873–1896.

The key summary statistics are direct. In Britain, the price index declines from 18.0 in 1873 to 14.7 in 1896. This corresponds to a circulation price change of -18.3%. The fixed-claim reservation index rises to 1.224, meaning that the real value of a fixed nominal claim increases by 22.4%. In the United States, the historical CPI declines from 36.0 in 1873 to 25.0 in 1896. This corresponds to a circulation price change of -30.6%. The fixed-claim reservation index rises to 1.440, meaning that the real value of a fixed nominal claim increases by 44.0%.

These values are not produced by a fitted model. They are arithmetic consequences of the historical price-level decline. Their theoretical significance lies in the domain distinction. A conventional CPI framework sees only negative inflation. SCR sees two simultaneous movements: the price of circulating goods falls, and the real value of fixed nominal claims rises. The latter is reservation inflation. For future extensions using public debt, interest rates, money, or other UK macro-financial variables, we also archived the Bank of England Millennium of UK Macroeconomic Data. [Thomas and Dimsdale 2017]

# 5. Results

## 5.1. The gold-standard price level declined rather than inflated

The first empirical result is straightforward: during the 1873–1896 gold-standard window, the circulation price level did not rise. It fell. In Britain, the retail price index declined from 18.0 in 1873 to 14.7 in 1896, corresponding to a cumulative circulation price change of -18.3%. In the United States, the historical CPI declined from 36.0 to 25.0 over the same window, corresponding to a cumulative circulation price change of -30.6%. The Minneapolis Fed source reports the U.S. series with 1967=100 and notes that pre-1913 values should be considered historical estimates; nevertheless, the broad nineteenth-century deflationary movement is large enough to support the present domain-level comparison. [FED Minneapolis, 2026]

This result confirms that the selected historical window is not a circulation-inflation episode. It is the opposite. If inflation is defined only as an increase in the consumer or retail price level, then the period should be described as deflation. That conventional description is correct at the level of circulating goods and services. However, SCR does not define inflation solely by the price movement of circulating commodities. It distinguishes between the circulation domain and the reservation domain. The fall in the circulation price level is therefore not the endpoint of the analysis; it is the condition under which the reservation mechanism becomes visible.

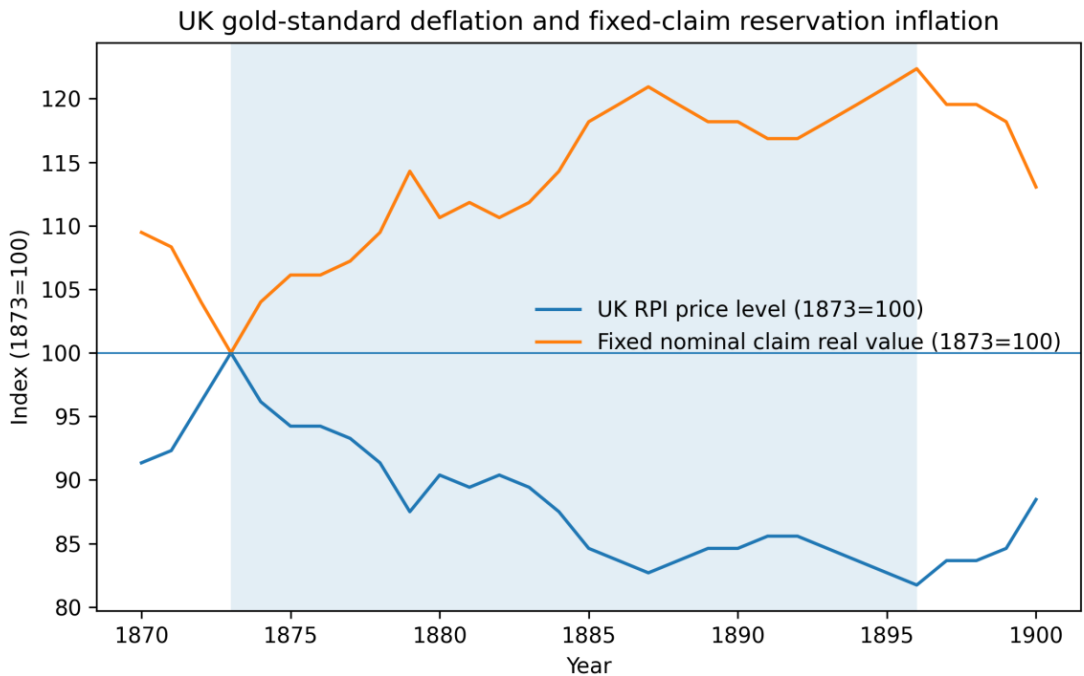

**Fig. 1. British circulation-price deflation and fixed-claim reservation inflation under the gold standard, 1870–1900.** The figure plots the British retail price index and the fixed-claim reservation inflation index over 1870–1900, with the 1873–1896 analysis window highlighted. The price index declines from 18.0 in 1873 to 14.7 in 1896, indicating negative circulation inflation. The fixed-claim reservation index, defined as $P_{1873}/P_t$, rises over the same period. The figure therefore shows the central SCR duality: the circulation price level falls while the real value of a fixed nominal claim rises. Data source: ONS long-run RPI series, CDKO.

The British case is especially important because Britain was central to the classical gold-standard system. The observed decline in the retail price index indicates that the gold-standard environment did not transmit monetary conditions into rising circulation prices. Instead, the price level adjusted downward. This is the empirical opposite of ordinary CPI inflation. Yet for a holder of a fixed nominal claim, the downward price adjustment increased real command over goods.

## 5.2. The transatlantic pattern was not a single-country artifact

The second result is that the British pattern is mirrored in the United States. When both price series are normalized to 1873=100, the British and U.S. price-level paths show the same qualitative movement: a sustained decline from the early 1870s to the mid-1890s. The magnitude is stronger in the U.S. series, but the sign is the same. Britain records an 18.3% decline in the price index from 1873 to 1896, while the United States records a 30.6% decline.

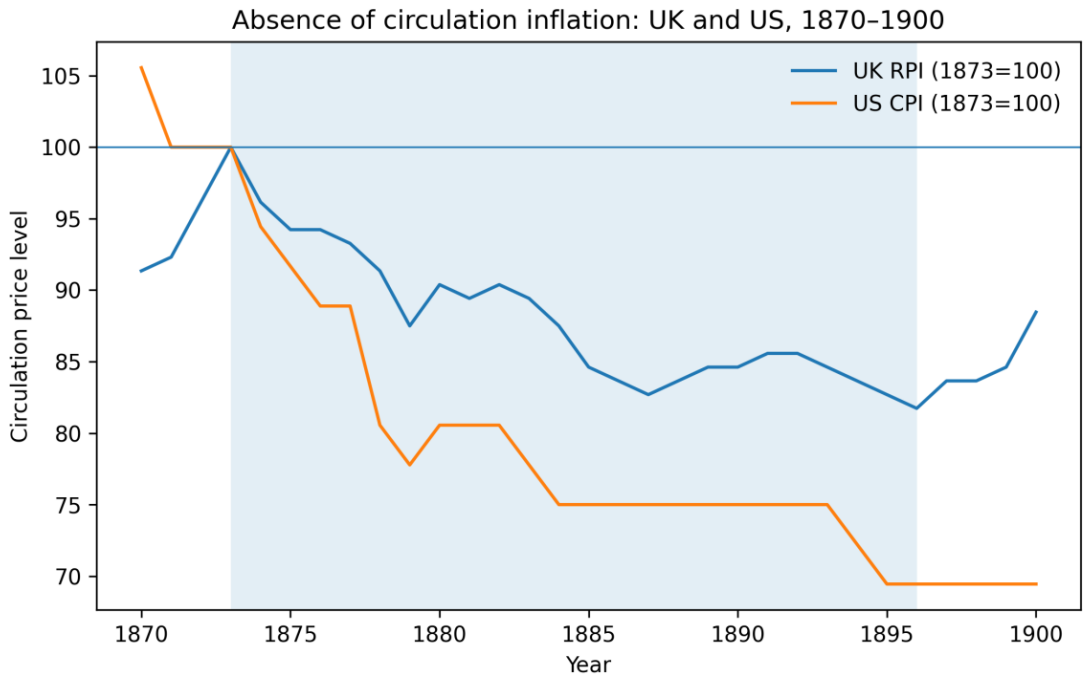

**Fig. 2. British and U.S. price levels normalized to 1873=100, 1870–1900.** The figure compares the British retail price index and the U.S. historical CPI after normalizing both series to 1873=100. Both countries show sustained price-level decline during the 1873–1896 window. This confirms that the analyzed episode was not a British idiosyncrasy, but part of a broader hard-money deflationary environment in the late nineteenth-century Atlantic economy. British data source: ONS CDKO. U.S. data source: Minneapolis Fed historical CPI table.

This cross-country comparison matters for the interpretation of the mechanism. If only one country showed price-level decline, the result could be attributed to local institutional or sectoral factors. The common downward movement in both Britain and the United States indicates that the episode reflects a broader monetary-material configuration: constrained monetary issuance, expanding production, and downward pressure on the circulation price level. The two economies were institutionally different, especially because the United States was still shaped by the monetary legacy of the Civil War and the return to gold convertibility [Friedman and Schwartz 1963], but the SCR mechanism does not require institutional identity.

It requires only the coexistence of falling circulation prices and durable nominal claims.

The U.S. comparison also strengthens the argument that the result is not a measurement artifact of one historical price index. The British and U.S. series have different origins, constructions, and institutional contexts, yet both point to the same domain-level result. Circulation inflation was absent; circulation deflation was present. The next question is therefore not whether prices rose, but what this price decline did to the real value of nominal claims.

### 5.3. Fixed nominal claims inflated in real terms

The third result is the central empirical test of SCR reservation inflation. For a fixed nominal claim, the reservation inflation index follows Eq. (6). It rises whenever the circulation price level falls. In Britain, the price decline from 18.0 to 14.7 implies that a fixed nominal claim increased its real value by 22.4% between 1873 and 1896. In the United States, the price decline from 36.0 to 25.0 implies that a fixed nominal claim increased its real value by 44.0%.

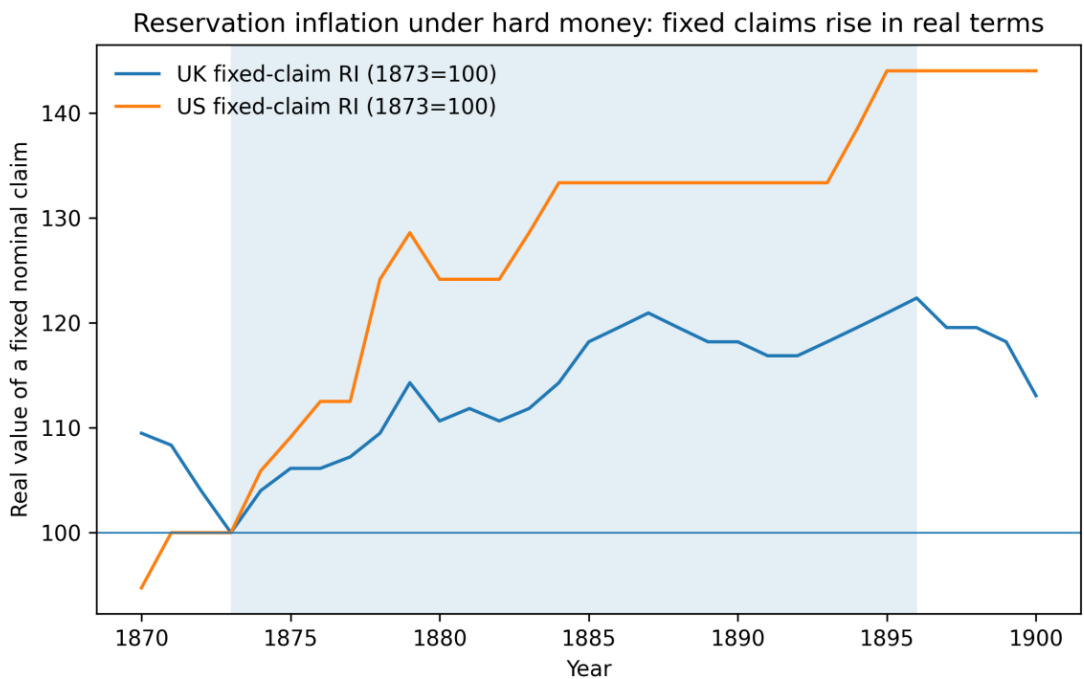

**Fig. 3. Fixed-claim reservation inflation index in Britain and the United States, 1870–1900.** The figure plots the real value of one unit of fixed nominal claim, normalized to 1873=1. The index is computed as $P_{1873}/P_t$. In both Britain and the United States, the index rises during the 1873–1896 window because the price level falls. By 1896, the British fixed-claim index reaches 1.224, while the U.S. fixed-claim index reaches 1.440. The result shows positive reservation inflation despite negative circulation inflation. British data source: ONS CDKO. U.S. data source: Minneapolis Fed historical CPI table.

This result is mechanical, but it is not trivial. Its importance lies in what it reveals about the meaning of deflation. In a conventional framework, the conclusion would be that the economy experienced negative inflation. In the SCR framework, the same data imply a dual process. The price of circulating goods fell, while the real weight of fixed nominal claims rose. A unit of money, a fixed debt contract, a bond face value, or another durable nominal claim obtained greater command over the physical goods represented by the price index.

This is the simplest empirical form of reservation inflation. It does not require broad asset-price appreciation, speculative bubbles, or expansionary central-bank balance sheets. It arises from the durability of nominal symbols under a falling circulation price level. The monetary claim does not need to increase in nominal size. It inflates in real terms because the goods against which it is measured become cheaper.

## 5.4. The SCR dual-inflation plane separates circulation deflation from reservation inflation

The fourth result summarizes the entire argument in a two-domain inflation plane. The horizontal axis measures cumulative circulation inflation from 1873 to 1896:

$$CI^{cum}_{1873 \to 1896} = \frac{P_{1896}}{P_{1873}} - 1$$

The vertical axis measures fixed-claim reservation inflation:

$$RI^{pct}_{1873 \to 1896} = \frac{P_{1873}}{P_{1896}} - 1$$

In this plane, both Britain and the United States fall in the quadrant defined by negative circulation inflation and positive reservation inflation. Britain is located at approximately (-18.3%, +22.4%). The United States is located at approximately (-30.6%, +44.0%).

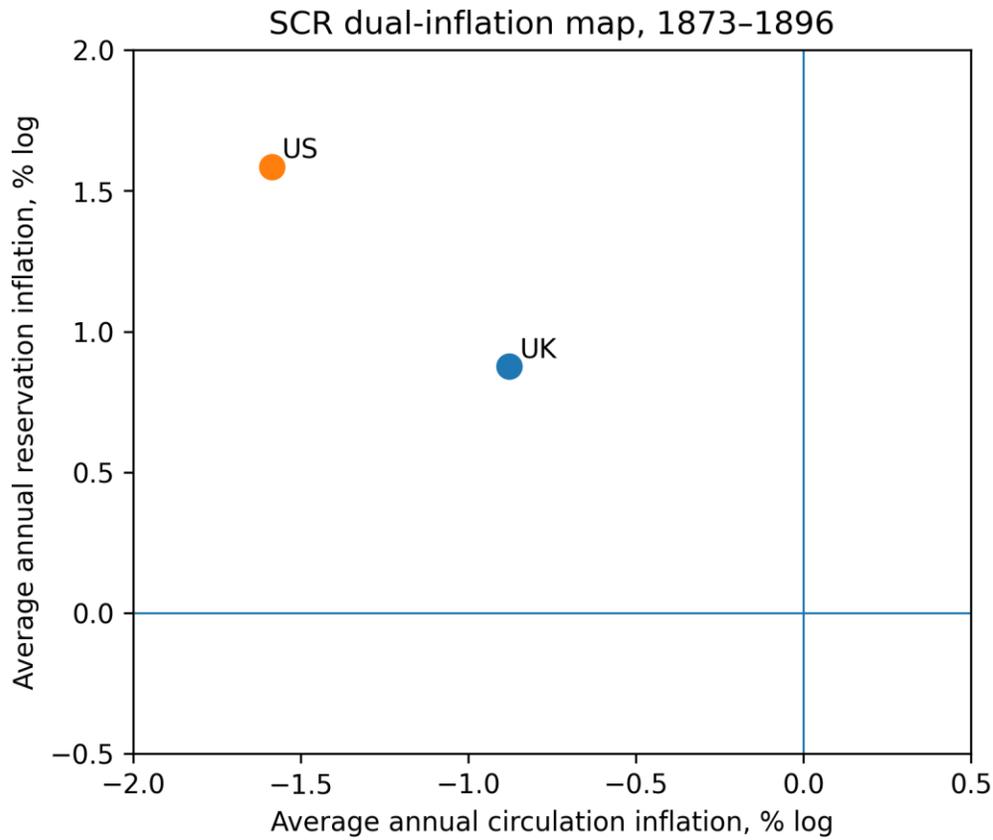

**Fig. 4. SCR dual-inflation map for Britain and the United States, 1873–1896.** The horizontal axis shows cumulative circulation inflation, following Eq. (10). The vertical axis shows fixed-claim reservation inflation, following Eq. (11). Britain and the United States both lie in the quadrant of negative circulation inflation and positive reservation inflation. This quadrant represents hard-money reservation inflation: the circulation price level falls, but the real value of fixed nominal claims rises. British data source: ONS CDKO. U.S. data source: Minneapolis Fed historical CPI table.

This figure is conceptually important because it prevents a category error. If the only axis is consumer-price inflation, the period appears anti-inflationary. Once the reservation axis is added, the same period becomes inflationary in a different domain. The two statements are not contradictory. They describe different objects. Circulation inflation asks whether goods prices rose. Reservation inflation asks whether monetary claims gained real weight relative to goods. In 1873–1896, the answer to the first question is no, while the answer to the second is yes.

The dual-inflation plane also clarifies the relationship between this paper and the Japan paper. The Japan case belongs to a different phase: monetary expansion was large, but CPI transmission weakened because money was absorbed into the reserve architecture. The gold-standard case is the opposite phase: monetary issuance was constrained, the circulation price level fell, and fixed nominal claims expanded in real value. Together, the two cases show that CPI alone is not a sufficient thermometer of monetary inflation. Inflationary pressure can disappear from the circulation domain while appearing in the reservation domain.

## 5.5. Endpoint arithmetic and cumulative interpretation

The endpoint arithmetic can be summarized in one compact table (Table 1).

**Table 1. Circulation-price change and fixed-claim reservation inflation, 1873–1896.**

| Country | Price index in 1873 | Price index in 1896 | Cumulative circulation inflation | Fixed-claim reservation inflation |
|---|---|---|---|---|
| Britain | 18.0 | 14.7 | -18.3% | +22.4% |
| United States | 36.0 | 25.0 | -30.6% | +44.0% |

Table 1 note: British price index values are from the ONS long-run RPI series, CDKO; U.S. CPI values are from the Minneapolis Fed historical CPI table; pre-1913 observations are historical estimates.

Table 1 makes the logic transparent. A decline in $P_t$ produces a negative circulation-inflation measure. The same decline produces a positive reservation-inflation measure because the real value of a fixed nominal claim is inversely related to the price level. The two columns therefore move in opposite directions by construction. This is not a statistical artifact; it is the core SCR identity.

The result should not be overinterpreted as a complete description of all nineteenth-century distributional changes. Actual historical outcomes depended on debt maturity, contract renegotiation, bankruptcy, wage flexibility, sectoral price differences, banking conditions, and political conflict. The fixed-claim index is a minimal theoretical instrument, not a full social accounting matrix. Its purpose is to isolate the hard-money reservation mechanism in its cleanest form. Once isolated, the mechanism is strong: a restrained monetary regime can eliminate circulation inflation while still increasing the real weight of nominal claims.

## 5.6. What the result proves, and what it does not prove

The empirical results prove a narrow but important point. They show that, in the 1873–1896 hard-money environment, the absence of circulation inflation did not imply the absence of inflationary redistribution in SCR terms. Fixed nominal claims appreciated in real value because the circulation price level fell. This is exactly the reservation-inflation side of SCR.

The results do not prove that all asset holders gained, that all debtors lost equally, or that every form of wealth behaved like a fixed nominal claim. They also do not claim that the gold standard was a purely fixed-money system. Credit conditions changed, gold supply changed, and banking systems evolved. The argument is more precise: whenever nominal claims remain durable while $P_t$ declines, those claims gain real value. The historical gold-standard deflation supplied such a setting at large scale.

This distinction protects the paper from an easy misunderstanding. The claim is not "the gold standard caused CPI inflation." It did not. The claim is not "deflation is secretly CPI inflation." It is not. The claim is that CPI deflation can coexist with, and mechanically generate, reservation inflation. In SCR terms, the gold standard did not abolish inflationary pressure. It moved the relevant inflationary expression from the circulation domain into the reservation domain.

# 6. Discussion

## 6.1. Deflation is not necessarily the opposite of inflation

The central implication of this paper is that deflation is not necessarily the opposite of inflation. It is only the opposite of circulation inflation. In conventional usage, inflation and deflation are defined by the sign of consumer-price change. A rising price level is inflation; a falling price level is deflation. That language is useful for measuring household purchasing costs, but it is incomplete as a theory of monetary-symbolic power. SCR adds a second domain: the reservation domain, where the relevant object is not the price of goods, but the real weight of nominal claims.

Under this two-domain framework, deflation can be inflationary in the reservation domain. When the price level falls, one unit of fixed nominal claim purchases more goods. The claim does not physically produce more goods, and it does not need to increase in nominal magnitude. Its real command expands because the denominator, $P_t$, declines. Thus, the same price movement that appears as deflation for circulating goods appears as inflation for monetary claims.

This result is especially important for hard-money systems. Hard money is often defended as anti-inflationary because it constrains overissue and protects the purchasing power of money. That description is partly correct. Hard money can suppress circulation inflation. But from an SCR perspective, the same mechanism can strengthen the reservation power of money. By preventing monetary symbols from depreciating alongside physical goods, hard money allows fixed nominal claims to gain real weight as productivity and supply expansion push prices downward.

The paradox is therefore only apparent. Hard money is anti-inflationary in the circulation domain and inflationary in the reservation domain. It constrains the price of goods from rising, but it permits the real value of monetary symbols to rise. It protects the holder of money and nominal claims, but it increases the burden on those who owe money or must obtain money through the sale of physical output.

## 6.2. The physical meaning of reservation inflation

The deeper SCR argument is physical rather than merely accounting-based. Physical goods exist in material time. They are consumed, worn out, replaced, stored at cost, or rendered obsolete. Monetary symbols are different. A nominal claim can persist as a ledger entry, bond face value, debt contract, or reserve balance without undergoing the same material decay. This asymmetry produces a structural imbalance between physical production and symbolic reservation.

In an expanding productive economy, goods may become more abundant and cheaper. If monetary symbols do not depreciate in parallel, then each unit of symbol commands more of the physical world. This is the physical core of hard-money reservation inflation. The monetary sign becomes increasingly powerful relative to the goods it claims, not because it has become more productive, but because the physical economy has lowered the circulation price of goods while the sign remains nominally intact.

This is why SCR treats inflation as a broader concept than consumer-price increase. The fundamental issue is not only whether prices rise. The issue is whether the relation between non-depreciating symbols and depreciating goods shifts. Under fiat overissue, the symbol may depreciate relative to goods, producing circulation inflation. Under hard-money deflation, goods depreciate in price relative to the symbol, producing reservation inflation. These are not opposites in the deepest sense. They are different phase expressions of the same symbolic-material asymmetry.

## 6.3. Relation to the Japan reserve-phase case

The present paper complements the Japan paper by testing the other side of SCR. The Japan case examined a modern fiat-reserve regime. [Huang 2026] After money entered a reserve-dominant phase, large expansions of the monetary base did not generate strong circulation inflation. The key empirical lesson was that monetary expansion can be absorbed into reserve structures rather than transmitted into consumer prices. That result supports the circulation side of SCR: CPI inflation is phase-dependent and may collapse when money changes phase.

The gold-standard case examines the opposite regime. Here, monetary issuance was constrained rather than expansive. Circulation inflation was absent not because money was trapped in modern reserve balances, but because the price level declined under a hard-money environment. Yet this absence of circulation inflation did not eliminate inflationary redistribution. It caused fixed nominal claims to rise in real value.

Together, the two cases form a symmetric pair. In Japan, high monetary expansion did not produce strong circulation inflation because money was absorbed into the reservation architecture. In the gold-standard case, low or constrained monetary expansion did not prevent reservation inflation because falling prices increased the real value of fixed claims. The common conclusion is that inflation cannot be identified with CPI alone. It must be located by phase and by domain.

This gives SCR a stronger empirical structure. The theory does not merely say that "money causes inflation" or that "money sometimes fails to cause inflation." It says that monetary asymmetry can appear in different domains depending on the institutional phase of money. In circulation-dominant regimes, it may appear as consumer-price inflation. In reserve-dominant fiat regimes, it may appear as balance-sheet and reserve absorption. In hard-money regimes, it may appear as the real expansion of fixed nominal claims.

## 6.4. Why this is not simply Fisherian debt deflation

The argument is related to, but not identical with, the traditional debt-deflation mechanism. [Fisher 1933] In debt-deflation theory, falling prices increase the real burden of debt, weakening borrowers and potentially triggering liquidation, distress, and financial contraction. SCR accepts this mechanism but places it within a broader theory of reservation inflation. Debt is one type of nominal claim, but not the only one. Money balances, bonds, rent claims, reserve claims, and other fixed nominal instruments also gain real weight when the price level falls.

The focus of SCR is therefore not only financial instability. It is the structural expansion of symbolic claims relative to physical goods. Debt-deflation theory emphasizes the crisis dynamics that can follow when real debt burdens rise. SCR emphasizes the prior domain transformation: before crisis occurs, the real value of nominal claims has already inflated. The debt-deflation spiral is a possible pathological consequence of reservation inflation, not the whole concept.

This distinction is useful because the 1873–1896 period was not a single acute collapse like the early 1930s. It was a long, relatively gradual deflationary environment. The reservation-inflation mechanism can operate in such a setting without requiring a sudden depression-scale crisis. That is why this paper uses 1873–1896 as the primary case rather than the Great Depression. [Bernanke James 1991] [Bernanke 1995] The latter is an extreme amplification of the mechanism; the former is a cleaner demonstration of its normal hard-money form. [Temin 1991]

## 6.5. Limits of the minimal fixed-claim index

The fixed-claim reservation index is intentionally minimal. It assumes a nominal claim that remains fixed over time. This is suitable for isolating the core mechanism, but it does not capture the full complexity of historical claims. Actual debts mature, roll over, default, renegotiate, or change interest rates. Bond prices fluctuate. Wages adjust. Real estate and equity claims are not fixed nominal claims. Sectoral price indices diverge. The economy is not a single representative commodity basket.

These limitations do not invalidate the result. They define its scope. The paper does not claim that every actor holding any asset experienced the same reservation inflation. It claims that the real value of fixed nominal claims rises mechanically when the price level falls. This is a theoretical baseline. More complex historical instruments can be studied as deviations from that baseline.

Future work can replace the fixed nominal claim assumption in Eq. (6) with observed historical series for public debt, private debt, bank deposits, consols, or government securities, thereby returning to the more general reservation index in Eq. (5). [Schularick and Taylor 2012] This special case is not a weakness. It is the cleanest way to show that reservation inflation can exist even without nominal claim expansion. [Capie and Webber 1985] If actual nominal claims also increase, then reservation inflation is stronger. If nominal claims contract, then the net effect depends on whether that contraction offsets the price-level decline.

## 6.6. The political economy of hard-money reservation inflation

Hard-money reservation inflation has a clear distributional implication. It favors holders of money and fixed nominal claims over debtors and producers whose income depends on selling goods at falling prices. When prices decline, a debtor must deliver more real goods or labor to satisfy the same nominal obligation. A creditor receives greater real value from the same nominal claim. Thus, hard-money deflation is not neutral. It redistributes real command over the physical economy toward the reservation side.

This helps explain why gold-standard deflation was politically contested. Debtors, farmers, producers, and expansionary monetary interests often opposed hard-money constraints because falling prices increased their real burdens. [Rockoff 1990] Creditors and holders of monetary claims often favored hard money because it preserved or increased the real value of their claims. SCR does not reduce this conflict to class rhetoric; it gives it a structural explanation. The conflict arises from the divergent movement of the circulation and reservation domains.

In this sense, reservation inflation is not merely a technical index. It measures a shift in command over the material economy. A rising fixed-claim reservation index means that the same monetary symbol can claim more goods. In physical terms, symbolic ownership gains weight relative to physical production. This is the deeper political economy of hard money.

## 6.7. Generalization beyond the gold standard

Although the gold standard provides a clean historical test, the mechanism is more general. Any regime with durable nominal claims and falling circulation prices can produce reservation inflation. This can occur under commodity money, currency boards, tight monetary policy, post-crisis deleveraging, technologically driven deflation, or productivity growth that outpaces monetary accommodation. The institutional details differ, but the general condition is still given by Eq. (3). If $d\ln P_t/dt$ is negative and $d\ln N_t/dt$ is not sufficiently negative to offset it, reservation inflation is positive. This condition is broad. It means that reservation inflation is not a curiosity of the nineteenth century. It is a general possibility whenever nominal claims are more durable than the price level of goods.

Modern economies usually avoid sustained consumer-price deflation through fiat monetary expansion, inflation targeting, and lender-of-last-resort institutions. [Bernanke 1995] But this does not abolish reservation inflation. It changes its form. Instead of fixed claims inflating through falling prices, reservation inflation may occur through balance-sheet expansion, asset-price inflation, reserve accumulation, public-debt absorption, or financial collateralization. The Japan paper addressed one such modern form. The gold-standard paper addresses the hard-money form.

## 6.8. Main implication for SCR

The main implication is that SCR cannot be tested by asking only whether CPI rises. [Huang 2018] The correct question is: in which domain does the monetary-material asymmetry appear? Japan shows that massive monetary expansion can fail to raise CPI when money enters a reserve-dominant phase. The gold-standard case shows that even constrained money can produce reservation inflation when prices fall. The common structure is phase displacement.

This result completes the empirical symmetry of the SCR program. The first side is that circulation inflation can disappear under reserve absorption. The second side is that reservation inflation can appear under hard-money deflation. Together, they support the broader claim that inflation is not a single scalar measured by CPI. It is a phase-dependent redistribution of symbolic command over physical goods.

# 7. Conclusion

This paper tested the reservation-inflation side of SCR. The original SCR theory argued that inflation should not be reduced to consumer-price inflation alone, but should be understood as a two-domain process involving both circulation inflation and reservation inflation. [Huang 2018] The Japan paper tested one side of this framework by showing that, in a reserve-dominant monetary phase, large monetary-base expansion may fail to transmit into strong consumer-price inflation. [Huang 2026] The present paper examined the opposite monetary condition: a hard-money regime in which issuance was constrained, circulation inflation was absent, and the price level declined.

The classical gold-standard deflation of 1873–1896 provides a clean historical test. In Britain, the price index declined from 18.0 in 1873 to 14.7 in 1896. In the United States, the historical CPI declined from 36.0 to 25.0 over the same period. These movements represent negative circulation inflation. Yet the same price decline mechanically increased the real value of fixed nominal claims. A fixed claim unchanged in nominal terms gained 22.4% in real value in Britain and 44.0% in the United States. Thus, the absence of consumer-price inflation did not imply the absence of inflationary redistribution. It implied a shift of inflationary expression from the circulation domain to the reservation domain.

This result changes the interpretation of hard money within SCR. A hard-money system may suppress circulation inflation, but it does not necessarily abolish inflationary pressure. By allowing the circulation price level to fall while nominal claims remain durable, it increases the real command of those claims over physical goods. In this sense, gold-standard deflation was not merely anti-inflationary. It was anti-inflationary for circulating goods, but inflationary for the real weight of fixed nominal claims.

Together with the Japan case, the present paper gives SCR a symmetric empirical structure. Japan shows that monetary expansion can lose its coupling to CPI when money enters a reserve-dominant phase. The gold-standard case shows that constrained money can still generate reservation inflation when prices fall. The two cases therefore support the same deeper principle: inflation is phase-dependent and domain-dependent. CPI is only one observable expression of the monetary-material asymmetry; in other regimes, the same asymmetry appears as reserve absorption, balance-sheet expansion, or the real appreciation of fixed nominal claims.

# Acknowledgment

This work is financially supported by the YRI-FD Industrial Project (YRI-IP-25-01).